\documentclass[english,aps,manuscript,prl,reprint]{revtex4-1}
\usepackage[T1]{fontenc}
\usepackage[latin9]{inputenc}
\usepackage{amsmath}
\usepackage{amssymb}
\usepackage{graphicx}

\makeatletter

\providecommand{\tabularnewline}{\\}

 \@ifundefined{textcolor}{}
 {%
   \definecolor{BLACK}{gray}{0}
   \definecolor{WHITE}{gray}{1}
   \definecolor{RED}{rgb}{1,0,0}
   \definecolor{GREEN}{rgb}{0,1,0}
   \definecolor{BLUE}{rgb}{0,0,1}
   \definecolor{CYAN}{cmyk}{1,0,0,0}
   \definecolor{MAGENTA}{cmyk}{0,1,0,0}
   \definecolor{YELLOW}{cmyk}{0,0,1,0}
 }

\usepackage{babel}

\usepackage{babel}

\usepackage{babel}

\usepackage{babel}

\usepackage{babel}

\usepackage{babel}

\usepackage{babel}

\makeatother

\usepackage{babel}
\begin{document}

\title{Simulating weak localization using a superconducting quantum circuit}

\author{Yu Chen$^{1}$}

\author{P. Roushan$^{1}$}

\author{D. Sank$^{1}$}

\author{C. Neill$^{1}$}

\author{Erik Lucero$^{1}$}

\altaffiliation[Present address: ]{HRL Laboratories, LLC, Malibu, CA 90265, USA}

\author{Matteo Mariantoni$^{1,2}$}

\altaffiliation[Present address: ]{Institute for Quantum Computing and Department of Physics and Astronomy, University of Waterloo, Waterloo, Ontario, Canada N2L 3G1}

\author{R. Barends$^{1}$}

\author{B. Chiaro$^{1}$}

\author{J. Kelly$^{1}$}

\author{A. Megrant$^{1,3}$}

\author{J. Y. Mutus$^{1}$}

\author{P. J. J. O'Malley$^{1}$}

\author{A. Vainsencher$^{1}$}

\author{J. Wenner$^{1}$}

\author{T. C. White$^{1}$}

\author{Yi Yin$^{1}$}

\altaffiliation[Present address: ]{Department of Physics, Zhejiang University, Hangzhou 310027, China}

\author{A. N. Cleland$^{1,2}$}

\author{John M. Martinis$^{1,2}$}

\email{martinis@physics.ucsb.edu}

\affiliation{$^{1}$Department of Physics, University of California, Santa Barbara,
California 93106-9530, USA}

\affiliation{$^{2}$California NanoSystems Institute, University of California,
Santa Barbara, CA 93106-9530, USA}

\affiliation{$^{3}$Department of Materials, University of California, Santa Barbara,
California 93106-9530, USA}

\maketitle
\textbf{Understanding complex quantum matter presents a central challenge
in condensed matter physics. The difficulty lies in the exponential
scaling of the Hilbert space with the system size, making solutions
intractable for both analytical and conventional numerical methods.
As originally envisioned by Richard Feynman, this class of problems
can be tackled using controllable quantum simulators \cite{Feynman1982,Buluta2009}.
Despite many efforts, building an quantum emulator capable of solving
generic quantum problems remains an outstanding challenge, as this
involves controlling a large number of quantum elements \cite{Aspuru-Guzik2012,Blatt2012,Bloch2012}.
Here, employing a multi-element superconducting quantum circuit and
manipulating a single microwave photon, we demonstrate that we can
simulate the weak localization phenomenon observed in mesoscopic systems.
By engineering the control sequence in our emulator circuit, we are
also able to reproduce the well-known temperature dependence of weak
localization. Furthermore, we can use our circuit to continuously
tune the level of disorder, a parameter that is not readily accessible
in mesoscopic systems. By demonstrating a high level of control and
complexity, our experiment shows the potential for superconducting
quantum circuits to realize scalable quantum simulators.}

\begin{figure*}
\begin{centering}
\includegraphics{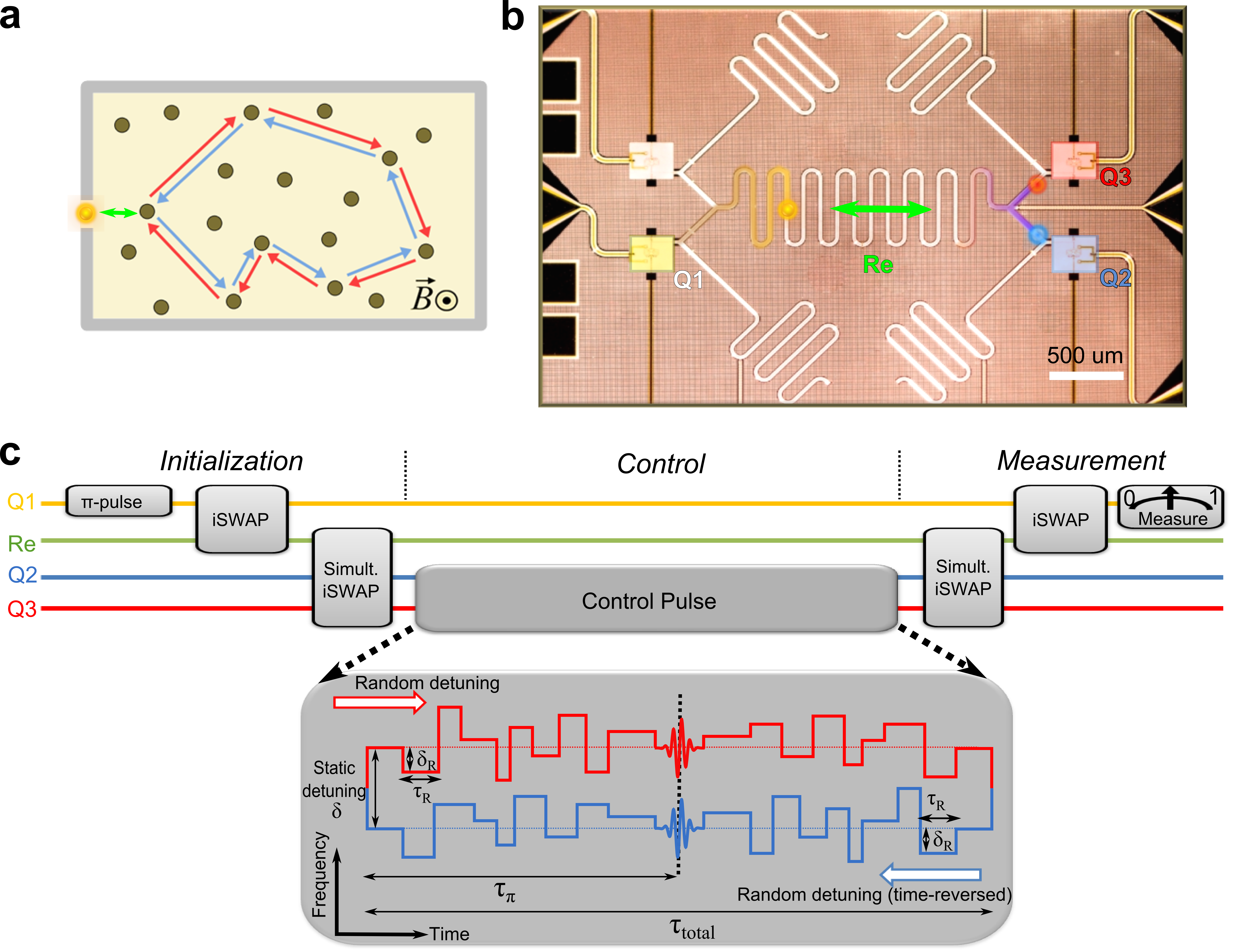} 
\par\end{centering}

\caption{\textbf{Superconducting quantum circuit implementation of weak localization}.
\textbf{a.} Schematic illustrating the basic physics of weak localization.
The weak localization contribution to electrical resistance is dominated
by the interference between closed trajectories traversed in opposite
directions, which in the small decoherence limit give an increase
in the resistance; here we show one such pair of trajectories. \textbf{b.}
Micrograph of the superconducting quantum circuit used to simulate
weak localization. A single microwave excitation is generated in qubit
$Q1$, distributed between $Q2$ and $Q3$ and the coupling resonator
$Re$, then manipulated in the simulation. \textbf{c.} Pulse sequence
used to simulate weak localization, decomposed into three steps: \textit{Initialization}:
A $\pi$-pulse creates an excitation in $Q1$ that is then swapped
into resonator $Re$ through an \textit{i}SWAP gate. The simultaneous
\textit{i}SWAP gate then transfers the excitation from $Re$ equally
to $Q2$ and $Q3$ through their three-body interaction, creating
a state $|\psi\rangle=|Q2\, Q3\, Re\rangle=\frac{1}{\sqrt{2}}(|eg0\rangle+|ge0\rangle)$
that entangles $Q2$ and $Q3$. \textit{Control:} We apply to $Q2$
a relative static detuning $\delta$, as well as a sequence of random
detunings $\delta_{R}$ each lasting for a random time $\tau_{R}$,
interspersed with a refocusing $\pi$-pulse spaced in time by $\tau_{\pi}$
to vary the effective coherence time $T_{\varphi^{\text{eff}}}$.
We apply the time-reversed sequence to $Q3$. \textit{Measurement}:
A simultaneous \textit{i}SWAP interferes the states of $Q2$ and $Q3$
in $Re$, and a second \textit{i}SWAP returns the excitation to $Q1$;
the probability for $Q1$ to be in the excited state $|e\rangle$
is then measured. }

\label{fig:concept} 
\end{figure*}

Superconducting quantum circuits have been used to simulate one- and
two-particle problems \cite{Neeley2009,Raftery2013}, and may be useful
for simulation of larger systems \cite{Houck2012}. Here, we use a
superconducting circuit to simulate the phenomenon of weak localization,
a mesoscopic effect that occurs in disordered electronic systems at
low temperatures. The challenge in this type of problem is that the
mesoscopic observables such as electrical resistance arise from the
interference of many scattering trajectories \cite{Bergmann1984},
thus apparently requiring a very large emulator. However, we find
that we can simulate weak localization using a time-domain ensemble
(TDE) approach: We sequentially run through many different circuit
parameter sets, each set simulating a different pair of scattering
trajectories in the mesoscopic system. By finding a one-to-one correspondence
between mesoscopic properties and quantum circuit parameters, we are
able to map the spatial complexity of the mesoscopic system onto a
set of complex yet manageable quantum control sequences in the time
domain.

\begin{table*}[t]
\begin{tabular}{c|c}
\hline 
\textit{Electron in mesoscopic system}  & \textit{Photon in quantum circuit}\tabularnewline
\hline 
\hline 
Magnetic field $B$  & Static detuning $\delta$\tabularnewline
Path area $S$  & Total detune pulse time $\tau_{\text{total}}$\tabularnewline
Wavevector $\overrightarrow{k}$  & Random detuning $\delta_{R}$\tabularnewline
Displacement $\overrightarrow{l}$  & Pulse duration $\tau$\tabularnewline
Coherence length $L_{\varphi}$  & Effective coherence time $T_{\varphi^{\text{eff}}}$\tabularnewline
Level of disorder  & Width of $\tau_{\text{total}}$ distribution $\sigma$\tabularnewline
\hline 
Electrical resistance $R$  & Photon return probability $P_{\text{return}}$\tabularnewline
\hline 
\end{tabular}

\caption{List of parameters for electron transport in a mesoscopic system,
and the corresponding control parameters in a quantum circuit.}

\label{table:parameters} 
\end{table*}

Weak localization involves the interference of electron trajectories
in a disordered medium \cite{Bergmann1984}. The quantum nature of
the electron allows it to simultaneously follow multiple trajectories,
each with amplitude $A_{n}$ and phase $\phi_{n}$. The probability
for the electron to reach a certain point is given by 
\begin{equation}
P=\sum_{n}A_{n}^{2}+\sum_{m\neq n}\sum_{n}A_{n}A_{m}\cos(\phi_{n}-\phi_{m}),
\end{equation}
where the first term sums over classical probabilities and the second
represents quantum interference. The quantum term typically averages
to zero, as scattering events randomize the electron wavevector $\overrightarrow{k}$
and displacement $\overrightarrow{l}$ and thus the accumulated phase
$\phi=\underset{j}{\sum}\,\overrightarrow{k_{j}}\cdot\overrightarrow{l_{j}}$.
A very dominant exception to this occurs in closed trajectories, as
these always have a time-reversed counterpart with identical accumulated
phase $\phi$ (Fig.\,\ref{fig:concept}(a)). These special pairs
thus interfere constructively with one another, yielding a probability
$P_{\text{return}}=4A^{2}$, twice the classical value. Experimentally,
weak localization is identified by applying a magnetic field $\overrightarrow{B}$,
which induces an additional static phase shift $\phi_{S}=2\left(2\pi\overrightarrow{B}\cdot\overrightarrow{S}\right)/\Phi_{0}$
for closed trajectories with area $\overrightarrow{S}$. The magnetic
field breaks the time-reversal symmetry and the precise constructive
interference of the paired closed trajectories. The measured electrical
resistance is thus maximum at zero applied field, and falls to the
classical resistance value as the magnetic field is increased - a
hallmark of weak localization. Reflecting quantum coherence from the
electron dynamics, weak localization is most known for its temperature
dependence. A elevated temperature increases the inelastic scattering
rates, reduces the phase coherence length $L_{\varphi}$, and thus
suppresses the magnitude of the weak localization peak at zero magnetic
field \cite{Bergmann1984,Pierre2003}.

\begin{figure*}
\begin{centering}
\includegraphics{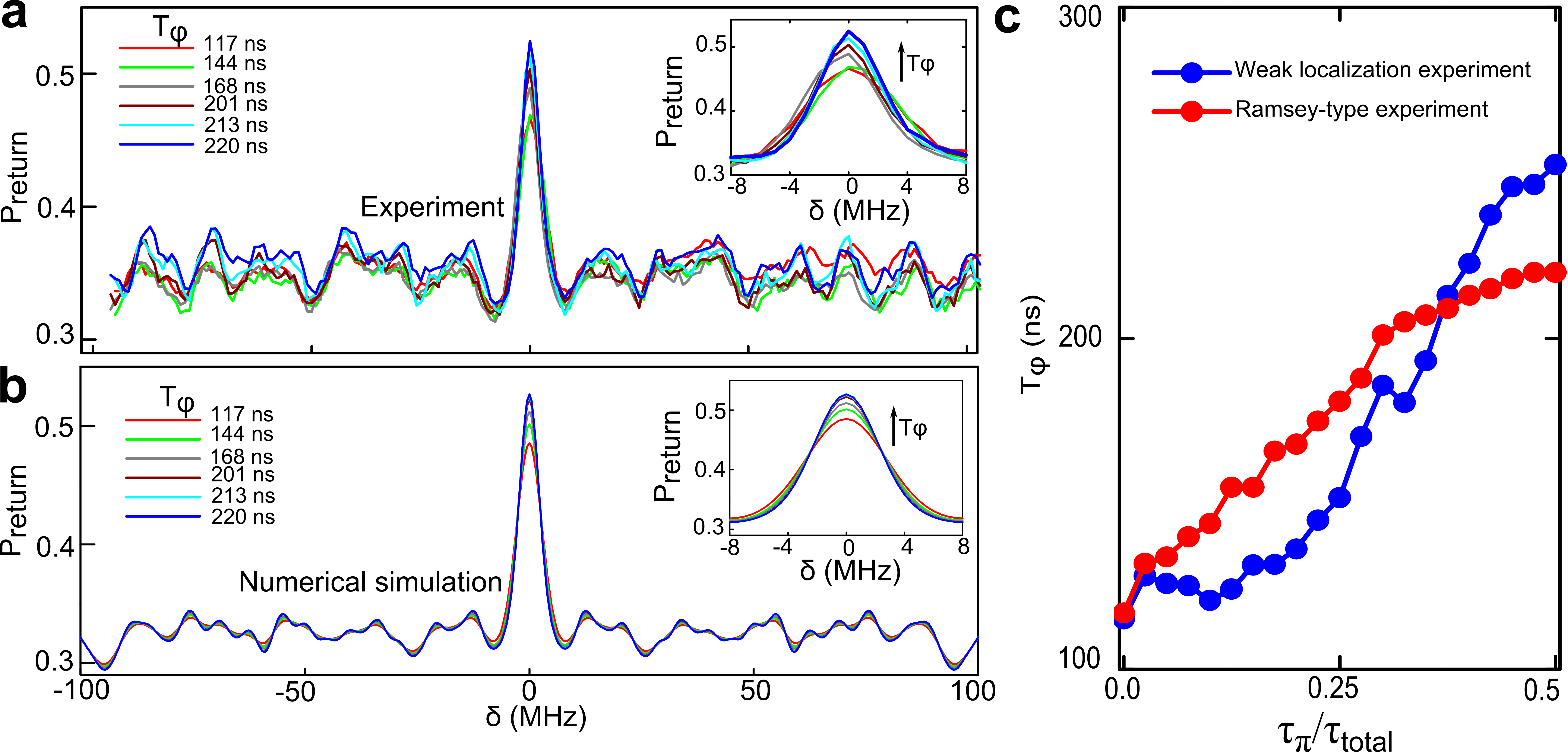} 
\par\end{centering}

\caption{\textbf{Simulating the temperature dependence of weak localization}
\textbf{a}. Measured photon return probability $P_{\text{return}}$
as a function of the static detuning $\delta$ (simulating a magnetic
field), for six different effective coherence times $T_{\varphi^{\text{eff}}}$.
The $P_{\text{return}}$ peak at zero detuning is analogous to the
magneto-resistance peak associated with weak localization. Inset shows
a magnified view of $P_{\text{return}}$ near $\delta=0$, where we
can observe the growth of the $P_{\text{return}}$ peak. This simulates
the growth of the magneto-resistance peak when lowering the temperature.
\textbf{b}. The photon return probability $P_{\text{return}}$ as
a function of $\delta$ obtained through numerical calculations, at
the same six different effective coherence time $T_{\varphi^{\text{eff}}}$
as the experiment. \textbf{c.} The effective coherence time $T_{\varphi^{\text{eff}}}$
extracted from the experiments, in comparison to $T_{\varphi^{\text{eff}}}$
directly measured by Ramsey-type experiments, as a function of $\tau_{\pi}/\tau_{\mbox{total}}$.}

\label{fig:temperature} 
\end{figure*}

Replacing the electron with a microwave excitation, we simulate weak
localization in a quantum circuit comprising three phase qubits, a
readout qubit \textit{Q1} and two control qubits \textit{Q2} and \textit{Q3},
symmetrically coupled to a bus resonator \textit{Re}, as shown in
Fig. \ref{fig:concept}b \cite{Lucero2012}. In this configuration,
the quantum circuit can be described by the Tavis-Cummings model\cite{Tavis1968}
\begin{equation}
H=\hbar\omega_{r}a^{+}a+\sum_{i=1}^{3}\hbar\omega_{i}\sigma_{i}^{+}\sigma_{i}^{-}+\sum_{i=1}^{3}\hbar g(a^{+}\sigma_{i}^{-}+a\sigma_{i}^{+}),
\end{equation}
where $\omega_{i}$ and $\omega_{r}$ are the frequencies of the qubits
and the resonator, respectively, and $g$ is the qubit-resonator coupling
strength.

As shown in Fig. \ref{fig:concept}(c), we start the simulation by
splitting the microwave excitation into the two control qubits, analogous
to an incoming electron simultaneously traversing two trajectories.
This was done by first initializing $Q1$ in $|e\rangle$, swapping
the excitation into $Re$ and then applying a simultaneous \textit{i}SWAP
gate \cite{Lucero2012,Mlynek2012}. By bringing $Q2$ and $Q3$ simultaneously
on resonance with $Re$ for an time $t=\pi/(2\sqrt{2}g)$, the simultaneous
\textit{i}SWAP gate transfers the excitation from $Re$ equally to
the two control qubits through their three-body interaction, resulting
in the desired state $|\psi\rangle=|Q2\, Q3\, Re\rangle=\frac{1}{\sqrt{2}}(|eg0\rangle+|ge0\rangle)$.

To simulate the the diffusion process of the electron in the presence
of magnetic field, we then apply a combination of sequences to the
control qubits, following the mapping between mesoscopic transport
and quantum circuit parameters delineated in Table \ref{table:parameters}.
To mimic the random scattering, we apply a series of random frequency
detunings $\delta_{R}^{i}$ to each qubit, each for a random duration
$\tau_{i}$ (see the extended pulse sequence in Fig.\ref{fig:concept}(c)),
resulting in a dynamic phase $\varphi_{R}=\underset{i}{\sum}\delta_{R}^{i}\cdot\tau_{R}^{i}$.
This simulates the random scattering phase $\phi=\underset{j}{\sum}\overrightarrow{k_{j}}.\overrightarrow{l_{j}}$
of the electron following a trajectory, with $\delta_{R}$ and $\tau_{R}$
corresponding to the electron wave vector $\overrightarrow{k}$ and
displacement $\overrightarrow{l}$, respectively. The random detuning
sequence applied to $Q3$ is the time-reversed sequence applied to
$Q2$, in order to properly simulate the time-reversal symmetry between
the direct and reversed electron trajectories. At the same time, we
apply a static detuning $\delta$ throughout the entire process, resulting
in a static phase $\varphi_{S}=\delta\cdot\tau_{\text{total}}$ between
the qubits. This simulates the magnetic field-induced phase shift
$\phi_{S}=2\left(2\pi\overrightarrow{B}\cdot\overrightarrow{S}\right)/\Phi_{0}$,
with $\delta$ and $\tau_{\text{total}}$ corresponding to $\overrightarrow{B}$
and $\overrightarrow{S}$, respectively.

To simulate the temperature dependence of weak localization, where
varying temperatures modifies the electron transport coherence length,
we insert a refocusing $\pi$-pulses into the sequence described above.
Instead of the conventional Hahn-echo sequence with the refocusing
pulse placed at $\tau_{\pi}/\tau_{\text{total}}=1/2$ \cite{Hahn1950},
we vary the timing of the refocusing pulse, such that the effective
phase coherence time can be continously modulated from $\sim100$\,ns
to over 200\,ns (measured by Ramsey-type experiments \textendash{}
see supplement). This projects the effective coherence time $T_{\varphi^{\text{eff}}}$
onto the electron phase coherence length $L_{\varphi}$.

Following this control sequence, we perform the measurement to the
system. We apply another simultaneous \textit{\small {{i}}}{\small {{SWAP}}}
gate, which allows the states of $Q2$ and $Q3$ interfere and recombine
to $Re$. At the end, an \textit{i}SWAP brings the interference result
back to $Q1$, with the probability of finding $Q1$ in $|e\rangle$
corresponding to the return probability of the electron in the direct
and reversed trajectories. Applying the TDE method discussed ealier,
we sequentially run through 100 different random detuning sequences
with different static and random detuning configurations, and find
the average return probability $P_{\mbox{return}}$, which is the
simulated electrical resistance for the mesoscopic transport problem.

\begin{figure*}
\begin{centering}
\includegraphics{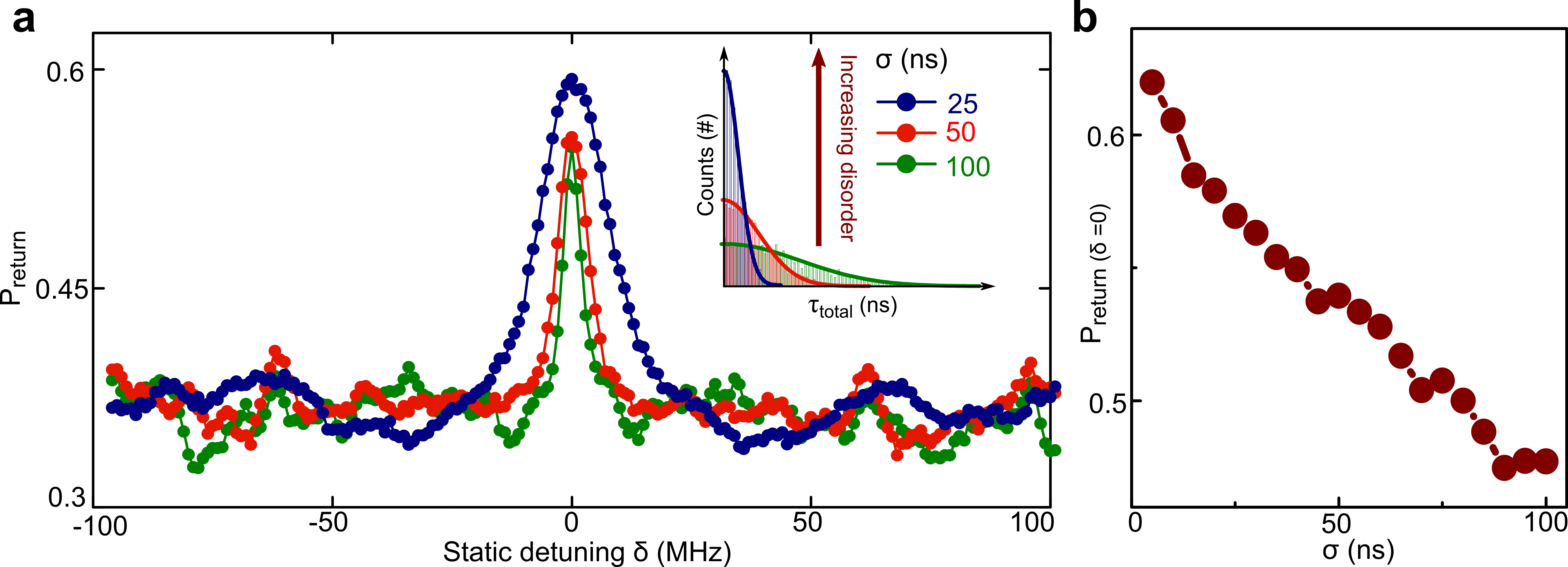} 
\par\end{centering}

\caption{\textbf{Simulating the disorder dependence of weak localization.}
\textbf{a}. Inset: Three distributions used for $\tau_{\text{total}}$
values, with standard deviations $\sigma=100$, 50 and 20 ns. Main
panel: $P_{\text{return}}$ as a function of $\delta$ for the distributions
shown in the inset. \textbf{b}. $P_{\text{return}}$ at zero detuning
as a function of the distribution width, showing the return probability
increasing as the distribution is reduced.}

\label{fig:disorder} 
\end{figure*}

In Fig. \ref{fig:temperature}(a), we show the experimental $P_{\text{return}}$
versus static detuning $\delta$ for six different $T_{\varphi^{\text{eff}}}$,
corresponding to magneto-resistance measurements at six different
temperatures. For all data sets, the probability $P_{\text{return}}$
has its maximum at $\delta=0$, where time-reversal symmetry is protected.
As $\delta$ moves away from zero, $P_{\text{return}}$ rapidly decreases
until it reaches an average value of approximately $0.35$, about
which it fluctuates randomly. The reduction in $P_{\text{return}}$
with increasing $|\delta|$ is consistent with the well-known negative
magneto-resistance in the mesoscopic system.

With the basic phenomena established, we focus on the small detuning
region to investigate the role of quantum coherence, through variations
in $T_{\varphi^{\text{eff}}}$ (Fig.\,\ref{fig:temperature}(a) inset).
While the overall structure remains unchanged, the $P_{\text{return}}$
peak grows as $T_{\varphi^{\text{eff}}}$ is increased; the peak rises
from $\sim0.46$ for $T_{\varphi^{\text{eff}}}=117$\,ns to $\sim0.53$
for $T_{\varphi^{\text{eff}}}=220$\,ns, consistent with the temperature
dependence of weak localization, where lower temperatures and thus
longer phase coherent lengths increase the magnitude of the negative
magneto-resistance peak \cite{Bishop1982}.

As they are performed on a highly controlled quantum system, our experimental
results can be understood within the Tavis-Cummings model (see supplement).
As shown in Fig. \ref{fig:temperature}b, we numerically evaluated
$P_{\text{return}}$ versus $\delta$, using the same six $T_{\varphi^{\text{eff}}}$
in the experiment. In the calculations we have also included the energy
dissipation time for each qubit, $T_{1}\sim500$\,ns. Except for
details in the aperiodic structures, the numerical results agree remarkably
well with our experimental observations.

Just as the phase coherence length $L_{\varphi}$ can be extracted
from magneto-resistance measurements displaying weak localization,
we can extract the effective coherence time $T_{\varphi^{\text{eff}}}$
from our measured $P_{\text{return}}(\delta)$. We measured $P_{\text{return}}(\delta=0)$
for various $\tau_{\pi}/\tau_{\text{total}}$, and subsequently extracted
$T_{\varphi^{\text{eff}}}$ from the height of the $P_{\text{return}}$
peak, based on the relationship $\Delta P_{\text{return}}\propto\frac{1}{2}\exp(-(\left\langle t\right\rangle /T_{\varphi^{\text{eff}}})^{2})\exp(-\left\langle t\right\rangle /T_{1})$
(see the theory section in supplement). The result is shown in Fig.
\ref{fig:temperature}c, compared with $T_{\varphi^{\text{eff}}}$
determined using conventional Ramsey-type measurements. As $\tau_{\pi}/\tau_{\text{total}}$
increases from 0 to 0.5, $T_{\varphi^{\text{eff}}}$ increases as
expected due to the cancellation of the qubit frequency drifts. We
find reasonable agreement between the values of $T_{\varphi^{\text{eff}}}$
as measured with the two techniques, with deviations possibly caused
by the finite number of ensembles in the simulation.

The importance of the weak localization effect is not only because
it reveals quantum coherence in transport, but also because it is
a precursor to strong localization, also known as Anderson localization
\cite{Anderson1958}. In the strong disorder limit, quantum interference
completely halts carrier transmission, producing a disorder-driven
metal-to-insulator transition. Our quantum circuit simulator allows
us to directly and separately tune the level of disorder, by varying
the distribution of pulse durations $\tau_{\mbox{total}}$, in contrast
with the mesoscopic system where tuning the disorder level typically
changes other parameters such as carrier density \cite{Lin1993,Ishiguro1992,Niimi2010}.

To measure the return probability $P_{\text{return}}$ as a function
of disorder, we average 100 random detuning pulse sequences with $\tau_{\text{total}}$
, where $\tau_{\text{total}}$ is randomly generated from a Gaussian
distribution with width $\sigma$. We used a Gaussian distribution
to mimic the diffusive nature of electron transport. The electron
displacement at a given time has a Gaussian distribution, where narrower
distributions correspond to greater disorder. Correlating the disorder
with $\sigma$, we simulate weak localization at increasing disorder
levels by reducing $\sigma$ from 100 to 50 to 25\,ns.

The experimentally measured $P_{\text{return}}$ versus $\delta$
for these three simulated disorder levels is shown in Fig. \ref{fig:disorder}a.
While the baseline value remains unchanged, the height of the zero-detuning
peak grows as we reduce $\sigma$. This growth in the peak height
with smaller $\sigma$ agrees with the observation that an increased
degree of disorder enhances localization in electron transport \cite{Lin1993,Ishiguro1992,Niimi2010}.

In order to find the signature of a disorder-driven metal-insulator
transition, we focus on $P_{\text{return}}$ at $\delta=0$ while
continuously reducing $\sigma$. The results, using $\tau_{\pi}/\tau_{\text{total}}=0.5$
for maximum $T_{\varphi^{\text{eff}}}$, are displayed in Fig. \ref{fig:disorder}b.
Reducing $\sigma$ results in an increase of the photon return probability,
with $P_{\text{return}}(\delta=0)$ increasing from 0.47 at $\sigma=200$
ns to 0.62 at $\sigma=10$ ns. However, there is no clear indication
of an abrupt transition to a fully localized state, which would correspond
to $P_{\text{return}}$ approaching unity. The metal-insulator transition
is therefore not observed in our current experiment. Observing this
transition likely requires further increasing the level of disorder,
i.e., increasing the ratio $T_{\varphi^{\text{eff}}}/\sigma$. Such
studies are now possible using the 100-fold improvement in coherence
time recently achieved using the Xmon qubit \cite{Barends2013}, and
are currently underway.

In closing, we comment on the aperiodic structure at the baseline
of $P_{\text{return}}$ that appears in both the experimental and
numerical results. The shape of this structure is independent of $T_{\varphi^{\text{eff}}}$,
while the fluctuation amplitude increases with increasing $T_{\varphi^{\text{eff}}}$.
These resemble the universal conductance fluctuations associated with
weak localization in mesoscopic systems: Both emerge from the frequency
beating of the interference fringes \cite{Umbach1984,Lee1985,Stone1985}.
Our experiment, however, does not include the cross-interference terms
between trajectories that do not have time-reversed symmetry, so it
is unclear if the fluctuation amplitudes here have a universal value
independent of the experimental details. Measurements on a quantum
system of large size are required to clarify this issue.

\section{Method}

The quantum circuit used in this experiment uses the same circuit
design as that used to implement Shor's algorithm \cite{Lucero2012}.
As shown in Fig.\ref{fig:concept}b, it is composed of four superconducting
phase qubits, each connected to a memory resonator and all symmetrically
coupled to a single central coupling resonator. The chip was fabricated
using conventional multi-layered lithography and reactive ion etching.
The different metal Al layers were deposited using DC sputtering and
the low-loss dielectric $a$-Si was deposited throught plasma-enhanced
chemical vapor deposition (PECVD).

The flux-biased phase qubit includes a $1$ pF parallel plate capacitor
and a 700 pH double-coiled inductor shunted with a Al/AlO{\small {{{x}}}}/Al
Josephson junction. The phase qubit can be modeled as a nonlinear
LC oscillator, whose nonlinearity arises from the Josephson junction.
Adjusting the flux applied to the qubit loop, we can modulate the
phase across the junction and consequently tune the qubit frequency.
We are thus able to tune the qubit frequency over more than several
hundred MHz without introducing any significant variation in the qubit
phase coherence. This property is crucial for this implementation
of the simulation protocol.

\section{Acknowledgments}

Devices were made at the UC Santa Barbara Nanofabrication Facility,
part of the NSF-funded National Nanotechnology Infrastructure Network.
This research was funded by the Office of the Director of National
Intelligence (ODNI), Intelligence Advanced Research Projects Activity
(IARPA), through Army Research Office grant W911NF-10-1-0334. All
statements of fact, opinion or conclusions contained herein are those
of the authors and should not be construed as representing the official
views or policies of IARPA, the ODNI, or the U.S. Government.

\section{Author contribution}

Y.C. conceived and carried out the experiments, and analyzed the data.
J.M.M. and A.N.C supervised the project, and co-wrote the manuscript
with Y.C. and P.R.. P.R., M.M, D.S. and C.N. provided assistance in
the data taking, data analysis and writing the manuscript. E.L. fabricated
the sample. All authors contributed to the fabrication process, qubit
design, experimental set-up and manuscript preparation.

\section{Competing financial interests}

The authors declare no competing finacial interests. 

\begin{thebibliography}{10}
\bibitem{Feynman1982} Feynman, R. Simulating physics with computers.
Int. J. Theor. Phys. 21, 467488 (1982).

\bibitem{Buluta2009} I. Buluta and F. Nori, Quantum simulators, Science
326, 108-111 (2009).

\bibitem{Aspuru-Guzik2012} A. Aspuru-Guzik and P. Walther, Photonic
quantum simulators, Nature Phys. 8, 285-291 (2012).

\bibitem{Blatt2012} R. Blatt and C. F. Roos, Quantum simulations
with trapped ions, Nature Phys. 8, 277-284 (2012).

\bibitem{Bloch2012} I. Bloch, J. Dalibard, and S. Nascimbene, Quantum
simulations with ultracold quantum gases, Nature Phys. 8, 267-276
(2012).

\bibitem{Neeley2009} M. Neeley, M. Ansmann, R. C. Bialczak, M. Hofheinz,
E. Lucero, A. D. O'Connell, D. Sank, H.Wang, J.Wenner, A. N. Cleland,
M. R. Geller, and J. M. Martinis, Emulation of a quantum spin with
a superconducting phase qudit, Science, 325, 722-725 (2009).

\bibitem{Raftery2013} J. Raftery, D. Sadri, S. Schmidt, H. E. T�reci,
and A. A. Houck, Observation of a dissipationinduced classical to
quantum transition, arXiv: 1312.2963 (2013).

\bibitem{Houck2012} A. A. Houck, H. E. Tureci, and J. Koch, On-chip
quantum simulation with superconducting circuits, Nature Phys. 8,
292-299 (2012).

\bibitem{Bergmann1984} G. Bergmann, Weak localization in thin lms
: a time-of-ight experiment with conduction electrons, Physics Reports
107, 1-58, 1984.

\bibitem{Pierre2003} F. Pierre, A. B. Gougam, A. Anthore, H. Pothier,
D. Esteve, and N. O. Birge, Dephasing of electrons in mesoscopic metal
wires, Phys. Rev. B 68, 085413 (2003).

\bibitem{Lucero2012} E. Lucero, R. Barends, Y. Chen, J. Kelly, M.
Mariantoni, A. Megrant, P. O'Malley, D. Sank, 10 A. Vainsencher, J.
Wenner, T. White, Y. Yin, A. N. Cleland, and J. M. Martinis, Computing
prime factors with a josephson phase qubit quantum processor, Nature
Phys. 8, 719-723 (2012).

\bibitem{Tavis1968} M. Tavis and F. W. Cummings, Exact solution for
an n-molecule-radiation-eld hamiltonian, Phys. Rev. 170, 379-384 (1968).

\bibitem{Mlynek2012} J. A. Mlynek, A. A. Abdumalikov, J. M. Fink,
L. Steen, M. Baur, C. Lang, A. F. van Loo, and A. Wallra, Demonstrating
w-type entanglement of dicke states in resonant cavity quantum electrodynamics,
Phys. Rev. A 86, 053838 (2012).

\bibitem{Hahn1950} E. L. Hahn, Spin echoes, Phys. Rev. 80, 580-594
(1950).

\bibitem{Bishop1982} D. J. Bishop, R. C. Dynes, and D. C. Tsui, Magnetoresistance
in si metal-oxide-semiconductor eld-eect transitors: Evidence of weak
localization and correlation, Phys. Rev. B 26, 773-779 (1982).

\bibitem{Anderson1958} P. W. Anderson, Absence of diusion in certain
random lattices, Phys. Rev. 109, 1492-1505 (1958).

\bibitem{Lin1993} J. J. Lin and C. Y. Wu, Electron-electron interaction
and weak-localization effects in ti-al alloys, Phys. Rev. B 48, 5021-5024
(1993).

\bibitem{Ishiguro1992} T. Ishiguro, H. Kaneko, Y. Nogami, H. Ishimoto,
H. Nishiyama, J. Tsukamoto, A. Takahashi, M. Yamaura, T. Hagiwara,
and K. Sato, Logarithmic temperature dependence of resistivity in
heavily doped conducting polymers at low temperature, Phys. Rev. Lett.
69, 660-663 (1992).

\bibitem{Niimi2010} Y. Niimi, Y. Baines, T. Capron, D. Mailly, F.-Y.
Lo, A. D. Wieck, T. Meunier, L. Saminadayar, and C. Bauerle, Quantum
coherence at low temperatures in mesoscopic systems: Effect of disorder,
Phys. Rev. B 81, 245306 (2010).

\bibitem{Barends2013} R. Barends, J. Kelly, A. Megrant, D. Sank,
E. Jerey, Y. Chen, Y. Yin, B. Chiaro, J. Mutus, C. Neill, P. O'Malley,
P. Roushan, J. Wenner, T. C. White, A. N. Cleland, and J. M. Martinis,
Coherent josephson qubit suitable for scalable quantum integrated
circuits, Phys. Rev. Lett. 111, 080502 (2013).

\bibitem{Umbach1984} C. P. Umbach, S. Washburn, R. B. Laibowitz,
and R. A. Webb, Magnetoresistance of small, quasi-one-dimensional,
normal-metal rings and lines, Phys. Rev. B 30, 4048-4051 (1984).

\bibitem{Lee1985} P. A. Lee and A. D. Stone, Universal conductance
uctuations in metals, Phys. Rev. Lett. 55, 1622-1625 (1985).

\bibitem{Stone1985} A. D. Stone, Magnetoresistance uctuations in
mesoscopic wires and rings, Phys. Rev. Lett. 54, 2692-2695 (1985).

\end{thebibliography}
\end{document}


\title{Supplementary Information for ``Simulating weak localization in
superconducting quantum circuit''}

\author{Yu Chen$^{1}$}

\author{P. Roushan$^{1}$}

\author{D. Sank$^{1}$}

\author{C. Neill$^{1}$}

\author{Erik Lucero$^{1}$}

\altaffiliation[Present address: ]{HRL Laboratories, LLC, Malibu, CA 90265, USA}

\author{Matteo Mariantoni$^{1,2}$}

\altaffiliation[Present address: ]{Institute for Quantum Computing and Department of Physics and Astronomy, University of Waterloo, Waterloo, Ontario, Canada N2L 3G1}

\author{R. Barends$^{1}$}

\author{B. Chiaro$^{1}$}

\author{J. Kelly$^{1}$}

\author{A. Megrant$^{1,3}$}

\author{J. Y. Mutus$^{1}$}

\author{P. J. J. O'Malley$^{1}$}

\author{A. Vainsencher$^{1}$}

\author{J. Wenner$^{1}$}

\author{T. C. White$^{1}$}

\author{Yi Yin$^{1}$}

\altaffiliation[Present address: ]{Department of Physics, Zhejiang University, Hangzhou 310027, China}

\author{A. N. Cleland$^{1,2}$}

\author{John M. Martinis$^{1,2}$}

\email{martinis@physics.ucsb.edu}

\affiliation{$^{1}$Department of Physics, University of California, Santa Barbara,
California 93106-9530, USA}

\affiliation{$^{2}$California NanoSystems Institute, University of California,
Santa Barbara, CA 93106-9530, USA}

\affiliation{$^{3}$Department of Materials, University of California, Santa Barbara,
California 93106-9530, USA}

\maketitle
\begin{figure}
\includegraphics{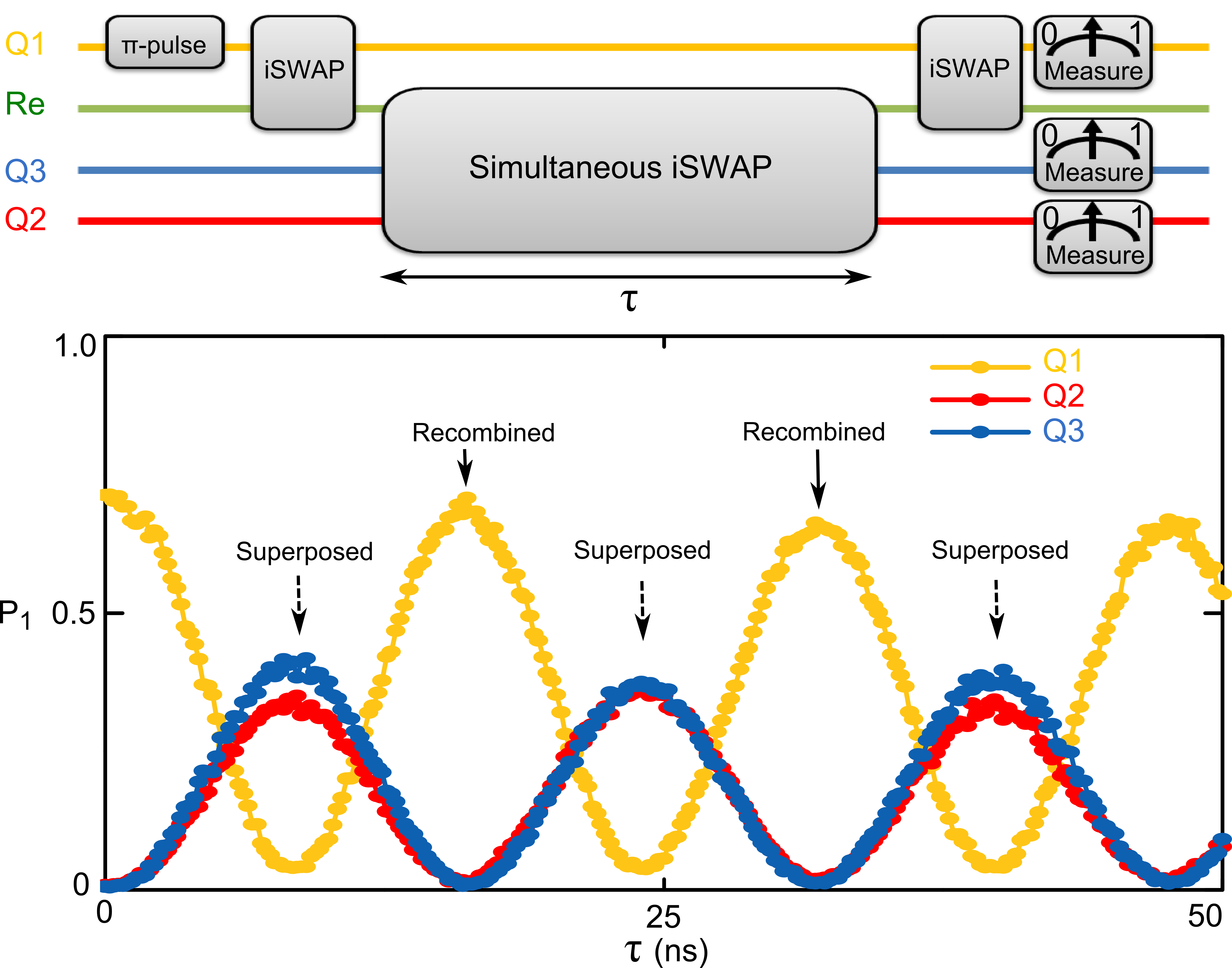}

\caption{\textbf{Upper.} The pulse sequence of distributing the photon between
the readout qubit and the two control qubits. \textbf{Lower.} The
distribution of the photon as a function of the control qubits-resonator
interaction time, where we can achieve the superposition and recombination
every $8.5$ ns. }

\label{fig:swap}
\end{figure}

\section{Photon distribution through the qubits-resonator collective interaction}

The simulation of weak localization requires the coherent photon transfer
between different quantum elements. In an architecture where all the
qubits are symmetrically coupled a center resonator, the quantum circuit
provides us a convenient way to coherently transfer a photon between
different elements, simply by tuning the qubits in and out of resonance
with the resonator.\cite{Mariantoni2011} As shown in the pulse sequence
in Fig. \ref{fig:swap}, we realize photon superposition and recombination
by distributing the photon between the readout qubit and the control
qubits, following the same protocol that has been employed to realize
W-type entangled state in superconducting quantum circuits.\cite{Lucero2012,Mlynek2012}
We first generated a photon in the readout qubit and have it transferred
to the coupling resonator. We then immediately detuned the readout
qubit back to its idling frequency, while bringing the two control
qubits on resonance with the coupling resonator. The two control qubits
then remain on resonance with the coupling resonator for a duration
$\tau$ , before we tuned them back to their original frequency and
have the remaining photon in the resonator transferred back into the
readout qubit. At the end, we perform measurements to all the three
qubits, determining the distribution of the photon. The result is
demonstrated in Fig. \ref{fig:swap}, where we plot the probability
of measuring the qubits to be in the excited state as a function of
the interaction time $\tau$. One can see that the photon is initially
concentrated in the readout qubit ($P_{1}$ maximum for Q1). After
an interaction time of $8.5$ ns, it splits evenly into the control
qubits ($P_{1}$ minimum for Q1). By maintaining the interaction for
the same duration time, we can reverse the process and have the photon
recombine back into the readout qubit.

\begin{figure}
\includegraphics{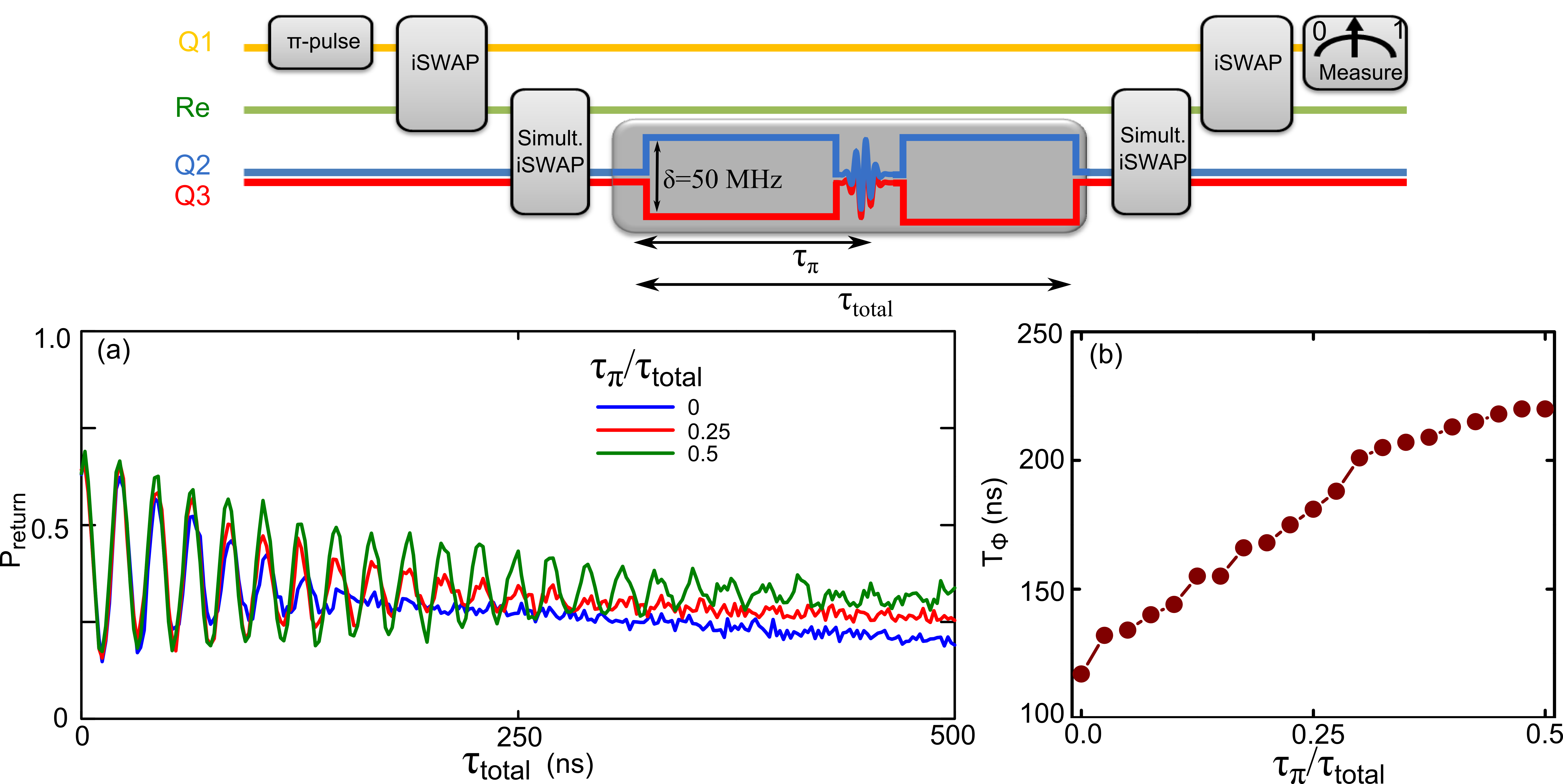}

\caption{\textbf{Upper.} Pulse sequence for the Ramsey-type interference experiment,
used to extract the overall coherence time $T_{\varphi}$ of the two
control qubits. \textbf{Lower.} a. P1 of Q1 as a function of $\tau_{total}$
obtained from the interference experiment, which we used to demonstrated
the tunability of the system phase coherence ,as we adjust the ratio
of $\tau_{\pi}/\tau_{total}$ . b. Extracted system coherence time
$T_{\varphi}$ as a function of $\tau_{\pi}/\tau_{total}$.}

\label{fig:coherence}
\end{figure}

\section{Tuning the system phase coherence time}

We simulated the temperature effect by tuning the phase coherence
time of our superconducting quantum system. One widely used method
to improve the system coherence time is the so-called Hahn-echo technique.\cite{Hahn1950}
By inserting a $\pi$-pulse into the middle of a pulse sequence, one
can refocus the phase of the qubit excitation and therefore effectively
compensate the system frequency drifting caused by 1/f flux noise.
In order to achieve a range of coherence times needed for simulating
different temperatures, we apply a modified Hahn-echo sequence, whose
effectiveness can be illustrated with a quantum interference experiment.
As shown by the pulse sequence in Fig. \ref{fig:coherence}, we first
prepare the photon in superposition state of occupying two control
qubits, following the method discussed in the previous section. We
then apply a constant detuning of $50$ MHz between the two control
qubits for a total time duration $\tau_{total}$, after which the
photon gets recombined and subsequently measured by the readout qubit.
Within the detuning pulse, we introduce refocusing $\pi$-pulses to
the two control qubits at a certain time $\tau_{\pi}$, used for effectively
tuning the system coherence time. As the results in Fig. \ref{fig:coherence}a
shows, when $\tau_{\pi}/\tau_{total}=0$ which corresponds to no refocusing
pulse, the interference fringes rapidly decay within the first $150$
ns. When $\tau_{\pi}/\tau_{total}=0.5$, which corresponds to the
standard Hahn-echo method, the interference fringes remains visible
even over $300$ ns, suggesting an improved phase coherence in the
system. A tunable coherence time between these two cases can therefore
be achieved by adjusting $\tau_{\pi}/\tau_{total}$ to have a value
between 0 and 0.5, with an example being demonstrated when $\tau_{\pi}/\tau_{total}=0.25$.
From the decay of the amplitude of the interference fringes, we extrapolates
the system effective coherence time $T_{\varphi^{\text{eff}}}$, which
basically averaged the dephasing rate of the two control qubits. As
shown in Fig. \ref{fig:coherence}b, as we move the location of the
refocusing pulse from the beginning to the middle of the detuning
sequence, the system effective phase coherence time $T_{\varphi^{\text{eff}}}$
gradually increases from its original value of $\sim117$ ns to $\sim220$
ns. These values are eventually used to simulate different temperatures
in the mesoscopic system.

\begin{figure}
\includegraphics{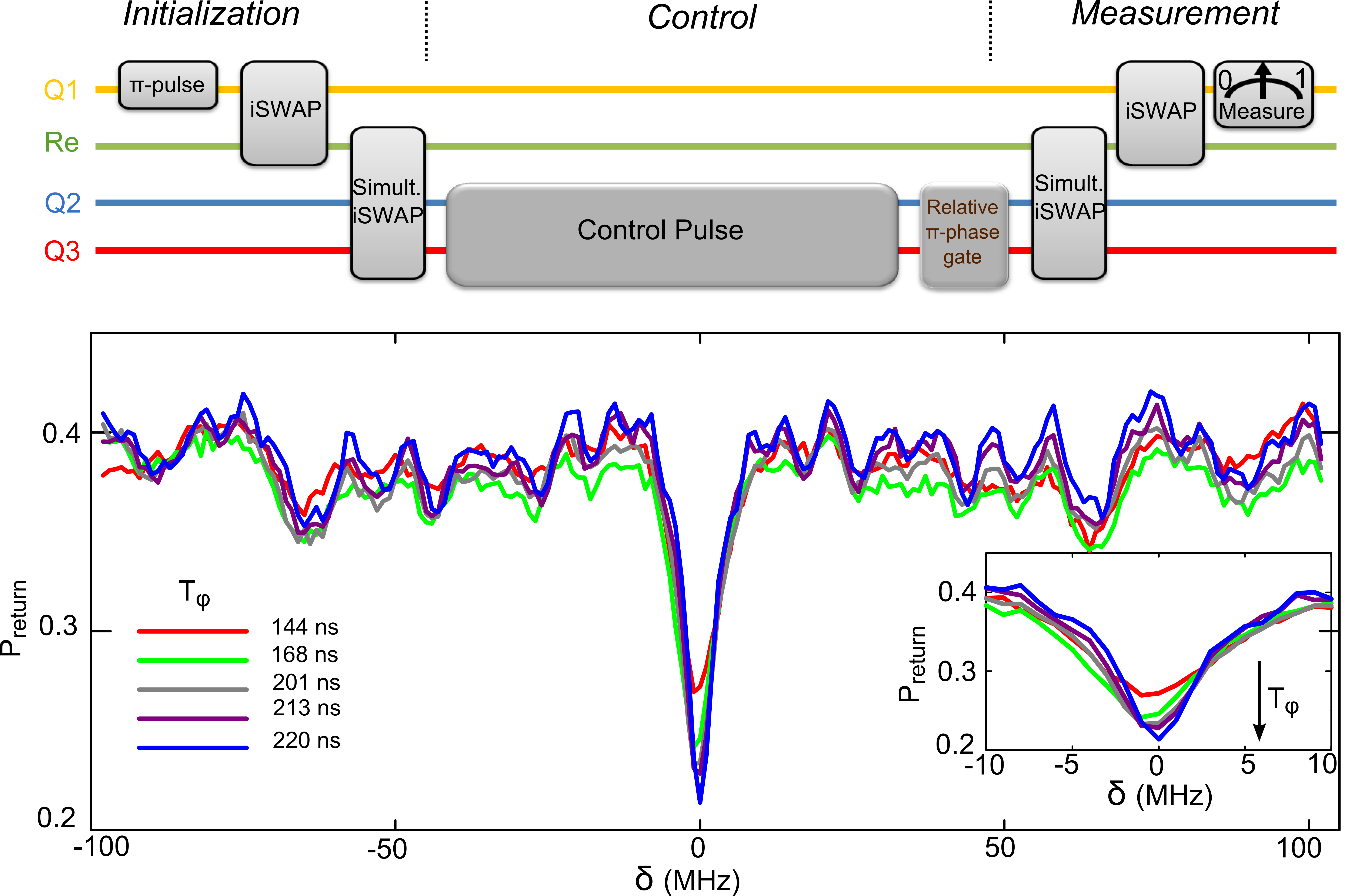}

\caption{\textbf{Upper.} Pulse sequence of simulating weak antilocalization,
where we append an additional detuning pulse to accumulate a relative
$\pi$ phase between splitted photon. \textbf{Lower.} The measured
photon return probability $P_{\text{{return}}}$ as a function of
the static detuning $\delta$, at different coherence time $T_{\varphi^{\text{eff}}}$.
The $P_{\text{{return}}}$ peak at zero detuning resembles the postive
magnetoresistance peak associated with weak antilocalization. Inset
shows a magnified view of $P_{\text{{return}}}$ near $\delta=0$,
where we can observe the growth of the $P_{\text{{return}}}$ valley,
simulating the growth of the magnetoresistance valley when lowering
the temperature.}

\label{fig:spin}
\end{figure}

\section{Weak anti-localization: Spin effect}

.

The existence of a sizable spin-orbit interaction in the mesoscopic
system has profound effects on the weak localization. In meoscopic
systems, the spin-orbit coupling induces momentum-dependent spin precession
during the scattering events, which shifts the spin phase oppositely
for the electron between time-reversed trajectories. The relative
phase shift inverts the original time-reversed symmetry into anti-symmetry,
which resultes in the well-known weak anti-localization.\cite{Yau2002,Bergmann1982,Hikami1980,Meir1989}
To simulate the effect of the added phase shift, we reprogram our
pulse sequences by appending an extra detuning pulse to the end of
the Control part of the pulse sequence, as shown in the top panel
of Fig. \ref{fig:spin}. The appended detuning pulse was precisely
calibrated to induce a relative $\pi$ phase rotation between the
splitted photon, which leads to the reversal of the symmetry between
the random detuning pulses. Running the reprogrammed sequence, we
remeasure $P_{\text{return}}$ as a function of the detuning $\delta$,
with the result demonstrated in Fig. \ref{fig:spin}. Under the symmetry
inversion, the photon now gains a higher probability to return as
we turned on the detuning $\delta$. The observation of the enhanced
photon return probability under the applied detuning $\delta$ corresponds
to the positive magneto-resistance in the mesoscopic system, the experimental
signature of the weak anti-localization. We further investigated the
temperature effect on the weak anti-localization by including the
Hahn-echo refocusing pulse into the sequence. As detailed data shown
in the inset of the figure, the simulated weak anti-localization gradually
loses its visibility as we tuned down the quantum coherence of the
system. This result is in agreement with well-established experimental
observations in mesoscopic systems, where an increased temperature
diminishes the amplitude of the positive magneto-resistance. \cite{Pierre2003}

\section{Theoretical discussion on photon distribution in Tavis-Cummings model}

Our quantum circuit, with all the qubit symmetrically coupled to the
bus resonator, can be described by the Tavis-Cummings model,
\begin{equation}
H=\hbar\omega_{r}a^{+}a+\sum_{i=0}^{3}\hbar\omega_{i}\sigma_{i}^{+}\sigma_{i}^{-}+\sum_{i=0}^{3}\hbar g(a^{+}\sigma_{i}^{-}+a\sigma_{i}^{+}),
\end{equation}

where $\omega_{r}$, $\omega$ are the resonate frequencies of the
resonance and qubits, and $g$ is the coupling strength between the
resonator and qubits. This allows for coherent photon transfer directly
between qubits and resonator, and indirectly from qubit to qubit via
the resonator.

In the case of the photon superposition and recombination, we brought
the two control qubits on resonance with the resonator while having
the other two qubits far detuned. In this case, we can obtain the
matrix for the Hamiltonian in the single photon subspace as
\begin{equation}
H_{1}=\left[\begin{array}{ccc}
\omega & g & g\\
g & \omega & 0\\
g & 0 & \omega
\end{array}\right].
\end{equation}
Diagonalizing the matrix, we can obtain the eigenenergies of the coupled
system as $E_{1}=\omega$, $E_{2}=\omega+g$ and $E_{3}=\omega-g$,
with three corresponding eigenstates $|\psi_{1}\rangle=\frac{1}{\sqrt{2}}(|0eg\rangle-|0ge\rangle)$,
$|\psi_{2}>=\frac{1}{\sqrt{2}}|1gg\rangle+\frac{1}{2}|0eg\rangle+\frac{1}{2}|0ge\rangle)$
and $|\psi_{3}\rangle=\frac{1}{\sqrt{2}}|1gg\rangle-\frac{1}{2}|0eg\rangle-\frac{1}{2}|0ge\rangle$.

For the photon superposition, we initialize the state as $|\psi(t=0)\rangle=|1gg\rangle$.
The time evolution of the wavefunction can be expressed as
\begin{equation}
|\psi(t)\rangle=\mbox{cos}\Omega t|1gg\rangle+\frac{i}{2}\mbox{sin}\Omega t|0eg\rangle+\frac{i}{2}\mbox{sin}\Omega t|0ge\rangle.
\end{equation}
We can see that the collective interaction allows the photon to oscillate
back and forth between the resonator and two control qubits, at a
frequency of $\Omega=\sqrt{2}g$.. If we set the interaction time
to be $t=(2n+1)\pi/(2\Omega)$, with n being an arbitrary integer
number, we can split the photon in the desired superposition state
$|\psi\rangle=\frac{1}{\sqrt{2}}(|0eg\rangle+|0ge\rangle)$.

For the photon recombination, the initial photon state is $|\psi(t=0)\rangle=\frac{1}{\sqrt{2}}(e^{i\phi_{1}}|0eg\rangle+e^{i\phi_{2}}|0ge\rangle)$,
where the phases $\phi_{1,2}$ contains both the random phase and
the static phase as described in the main text. Ignoring a global
phase and setting $\phi=\phi_{1}-\phi_{2}$, we obtain the time evolution
of the wavefunction as
\begin{align}
|\psi(t)\rangle & =\frac{i}{2}(e^{i\phi}+1)\cdot\mbox{sin}\Omega t|1gg\rangle\nonumber \\
 & +\frac{1}{2\sqrt{2}}[(1-e^{i\phi})+\mbox{cos}\Omega t\cdot(1+e^{i\phi})]|0eg\rangle\nonumber \\
 & +\frac{1}{2\sqrt{2}}[-(1-e^{i\phi})+\mbox{cos}\Omega t(1+e^{i\phi})]|0ge\rangle.
\end{align}

We set the interaction time to be $t=(2n+1)\pi/(2\Omega)$ for a photon
recombination. As a consequence, the probability for the photon to
be in the resonator, which is eventually measured by Q1, is,
\begin{equation}
P_{1}=\frac{1}{2}(1+\cos\phi).
\end{equation}

The photon return probability oscillates when we tune the relative
phase between two branches of the splitted photon, demonstrating a
microwave photon version of the Aharonov\textendash{}Bohm effect.

In the experiment, the relative phase $\phi$ was accumulated for
a certain detune time $t$. During this period of time, the fluctuations
in the frequency can lead to the dephasing of the qubits, exhibited
as a phase noise $\Delta\phi$. Averaged over these phase fluctuations,
Eq. (5) gets modified into

\begin{align}
P_{1} & =\left\langle \frac{1}{2}(1+\cos(\phi+\Delta\phi))\right\rangle \nonumber \\
 & =\frac{1}{2}(1+\left\langle \cos\phi\cos(\Delta\phi)\right\rangle )\nonumber \\
 & =\frac{1}{2}(1+\cos\phi\left\langle \cos(\Delta\phi)\right\rangle ),
\end{align}

where we have assumed both qubits to have the same dephasing rate,
$\left\langle \right\rangle $ stands for averging over random phase
fluctuations and in the last line we have used $\left\langle \sin(\Delta\phi)\right\rangle =0$,
taking into account of the even distribution of $\Delta\phi$.

As thoroughly discussed in Ref.\,\cite{Martinis2003}, dephasing
in this case can lead to the decay of the photon, given as
\begin{equation}
P_{1}=\frac{1}{2}(1+\cos\phi\cdot\exp(-(t/T_{\varphi1}+(t/T_{\varphi2})^{2})),
\end{equation}

where the exponential decay originated from the white noise and the
Gaussian decay originated from the $1/f$ noise.

Finally, taking into account for dissipation throughout the sequence,
we can obtain the full expression to find the photon back in the resonator
as
\begin{equation}
P_{1}=\frac{1}{2}(1+\cos\phi\cdot\exp(-(t/T_{\varphi1}+(t/T_{\varphi2})^{2}))\exp(-t/T_{1}).
\end{equation}

In the experiment, we modulate the effect of the $1/f$ noise by adjusting
the timing of the inserted a $\pi$-pulse in the control sequence.
In this way, we can effectively modulate $T_{\varphi2}$, which is
referred as $T_{\varphi^{\text{eff}}}$ as in the main text.